\documentclass[prd,tightenlines]{revtex4}
\usepackage{epsfig}

\usepackage{amssymb}
\usepackage{amsfonts}
\usepackage{graphicx}
\usepackage{keyval,graphicx}
\usepackage{textcomp,wasysym}
\begin{document}

\bibliographystyle{unsrt}

\title{A possible mechanism for producing the threshold enhancement in $J/\psi\to\gamma p\bar{p}$ }

\author{Xiao-Hai Liu$^1$, Yuan-Jiang Zhang$^1$ and Qiang Zhao$^{1,2}$}

\affiliation{1) Institute of High Energy Physics, Chinese Academy of
Sciences, Beijing 100049, P.R. China \\
2) Theoretical Physics Center for Science Facilities, CAS, Beijing
100049, P.R. China}

\date{\today}

\begin{abstract}
In the $J/\psi$ radiative decay channels $J/\psi\to \gamma
V\bar{V}$, the result of partial wave analysis indicates that the
$V\bar{V}$ systems are predominately pseudoscalar component, and
most of these channels have relatively large branching ratios at an
order of $10^{-3}$. Meanwhile, vector mesons, such as $\rho$,
$\omega$ and $K^*$, have strong couplings with nucleons and/or
hyperons. This suggests a dynamical mechanism describing the $\eta
p\bar{p}$ form factors for higher $\eta$ mesons, such as
$\eta(1405/1475)$ and $\eta(1760)$. It is thus natural to expect
that rescatterings of these vector meson pairs into $p\bar{p}$ of
$0^-$ partial wave could be an important source contributing to
$J/\psi\to\gamma p\bar{p}$ of which the branching ratio is at the
order of $10^{-4}$. Our calculation justifies this point. In
particular, we find that interferences between different
rescattering amplitudes can produce a significant threshold
enhancement in the invariant mass spectrum of $p\bar{p}$. Without
introducing dramatic ingredients, our model provides a natural
explanation for the peculiar threshold enhancement observed by
BES-II in $J/\psi\to \gamma p\bar{p}$. Additional experimental
constraints on the $V\bar{V}\to p\bar{p}$ transitions are examined.
This mechanism in $J/\psi\to \omega p\bar{p}$ is also discussed.
\vspace{0.5cm}


\end{abstract}

\maketitle

 %13.20.Gd Decays of <i>J</i>/psi, Upsilon, and other quarkonia
 %13.38.-b Decays of intermediate bosons
 %13.30.Eg Hadronic decays

\section{Introduction}
BES-II Collaboration has reported a narrow threshold enhancement
near $2m_p$ in the invariant mass spectrum of $p\bar{p}$ pairs from
$J/\psi\to \gamma p\bar{p}$ decays \cite{Bai:2003sw}. The result of
partial wave analysis (PWA) shows that if it is interpreted as a
$0^{-+}$ resonance, its mass is about
$M=1859^{+3}_{-10}(\mbox{stat})^{+5}_{-25}(\mbox{syst})\
\mbox{MeV}$, and its decay width is about $ \Gamma < 30\ \mbox{MeV}$
at $90\%$ C.L. This observed enhancement has stimulated many
theoretical studies of its underlying structure. Some interpret it
as a glueball candidate \cite{Kochelev:2005vd,Li:2005vd,Hao:2005hu}
or a baryonium
\cite{Datta:2003iy,Ding:2005ew,Loiseau:2005cv,Zhu:2005ns}, and some
others take into account the effect of  the final state $p\bar{p}$
interactions
\cite{Zou:2003zn,Kerbikov:2004gs,Bugg:2004rk,Sibirtsev:2004id,Entem:2007bb,Chen:2008ee,Haidenbauer:2008qv}.
There will be some peculiar characters if it is interpreted as a
glueball. For instance, it can couple to a pair of vector mesons
$V\bar{V}$, and the decay channel will be flavor independent, but we
have not yet found such a narrow state in $J/\psi\to \gamma
V\bar{V}$ decays considering there has been a sizeable accumulation
of events.

On the other hand, there exists an interesting phenomenon that may
be related to the $J/\psi\to\gamma p\bar{p}$ decay. We notice that
in the process of $J/\psi$ radiative decays $J/\psi\to\gamma
V\bar{V}$, where $V\bar{V}$ represent $\rho\rho$, $\omega\omega$ or
$K^*\bar{K}^*$ and so on, the $V\bar{V}$ invariant mass distribution
is dominated by the $0^-$ components in all of these channels
\cite{Bugg:1995jq,Ablikim:2006ca,Bai:1999mk,Baltrusaitis:1985nd,Bisello:1988as}.
Nevertheless, these vector mesons generally have strong couplings
with the nucleons and/or hyperons. As a result, we expect that the
$V\bar{V}$ rescattering into $p\bar{p}$ could play an important role
in the description of the pseudoscalar-$p\bar{p}$ coupling form
factor in $J/\psi\to\gamma p\bar{p}$. This is also consistent with
that the $p\bar{p}$ system has an important $0^-$ component. We list
in Tab.~\ref{table1} some relatively significant channels of which
the branching ratios are at the order of $10^{-3}$ and which might
contribute to the rescattering. It is worth noting that the
experimental values of $BR(J/\psi\to\gamma\eta(1405/1475))\times
BR(\eta(1405/1475)\to \rho\rho)$ and
$BR(J/\psi\to\gamma\eta(1760))\times BR(\eta(1760)\to \rho\rho)$ are
extracted from $J/\psi\to 4\pi\gamma$, where the results greatly
depend on the fitting methods \cite{Bisello:1988as}. A similar
problem is also with the data for $BR(J/\psi\to\gamma 0^-)\times
BR(0^-\to K^*\bar{K}^*)$~\cite{Bai:1999mk}. In this sense, the
branching ratios displayed in Table~\ref{table1} still have large
uncertainties at the order of $10^{-3}$. However, we shall show
later that the sizeable $V\bar{V}$ rescatterings into $p\bar{p}$
cannot be neglected at all if the uncertainties were not more than
one order of magnitude. For the purpose of exploring possibilities
of reproducing the line shape of the threshold enhancement in
$J/\psi\to\gamma p\bar{p}$~\cite{Bai:2003sw}, we can adopt such a
set of values to determine the coupling constants and examine the
model-dependent and independent aspects in this decay transition.

BES-II also reported another similar resonance observed in the
$\pi^+\pi^-\eta'$ invariant mass spectrum in $J/\psi\to
\gamma\pi^+\pi^-\eta'$, which has a mass $M=1833.7\pm
6.1(\mbox{stat})\pm 2.7(\mbox{syst})\ \mbox{MeV}$ and a width
$\Gamma=67.7\pm 20.3(\mbox{stat})\pm 7.7(\mbox{syst})\ \mbox{MeV}$
after a fit with a Breit-Wigner function \cite{Ablikim:2005um}. If
these two experimental results can be attributed to the same
resonance, it would be an additional evidence for the resonant
property of this enhancement. But we should note that the present
experimental data do not allow one to conclude whether
$X(1835)\to\pi^+\pi^-\eta'$ is via quasi-two-body decay (e.g.
through $X(1835)\to\sigma\eta'\to \pi^+\pi^-\eta'$) or three-body
decay. In order to understand the nature of the threshold
enhancement in $J/\psi\to \gamma p\bar{p}$, one should explore
various possibilities in the transition mechanism. This forms our
motivation in this work to study the role played by vector meson
$V\bar{V}$ rescatterings in $J/\psi\to \gamma p\bar{p}$.

\begin{center}
\begin{table}
\begin{tabular}{|c|c|c|}
  \hline
  \hline
  Channel & BR ($\times 10^{-3}$) & Ref. \\ \hline
  $\gamma\eta(1405/1475)\to\gamma\rho\rho$ & $1.83\pm 0.39$ & \cite{Bisello:1988as} \\
  $\gamma\eta(1760)\to\gamma\rho\rho$ & $1.44\pm 0.33$ & \cite{Bisello:1988as} \\
  $\gamma\eta(1760)\to\gamma\omega\omega$ & $1.98\pm 0.33$  & \cite{Ablikim:2006ca} \\
  $\gamma 0^{-}\to\gamma K^*\bar{K}^*$ & $2.3\pm 0.9$  & \cite{Bai:1999mk} \\
  \hline
\end{tabular}
\caption{Branching ratios of $J/\psi\to\gamma\eta\to\gamma
V\bar{V}$, where $0^-$ represents a broad $0^-$ resonance with the
mass $M=1800\pm 100\ \mbox{MeV}$ and the decay width $\Gamma=500\pm
200\ \mbox{MeV}$ \cite{Bai:1999mk}.} \label{table1}
\end{table}
\end{center}

As follows, we first provide details of our theoretical model in
Sect. II. Numerical results and discussions will be given in Sect.
III. A brief summary will be given in Sect. IV.

\section{The Model}
\begin{figure}[t]
\epsfig{file=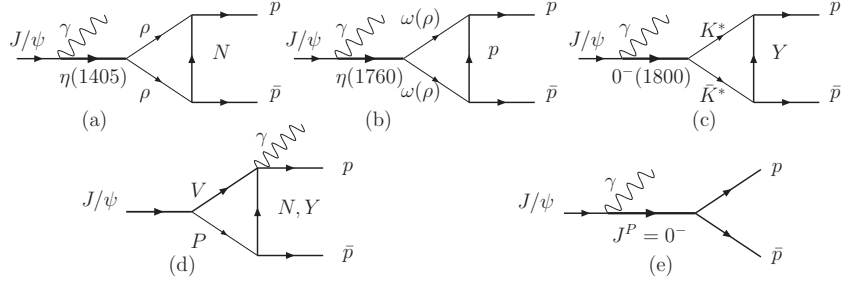,scale=0.55} \caption{Feynman diagrams for
$J/\psi\to\gamma p\bar{p}$. Diagrams (a)-(c) are via $V\bar{V}$
rescattering, where $N$ and $Y$ represent the exchanged nucleon or
hyperon, respectively. Diagram (d) is $VP$ rescattering, where $V$
and $P$ represent vector and pseudoscalar meson, respectively, such
as $\rho\pi$, $K^*\bar{K}$ and so on. Diagram (e) is $p\bar{p}$
production via direct couplings to pseudoscalar
resonances.}\label{fig1}
\end{figure}

The Feynman diagrams that illustrate the rescattering transitions
are shown in Fig.~\ref{fig1}. Considering that the coupling
constants of $K^*N\Sigma$ are smaller than those of $K^*N\Lambda$,
especially the tensor coupling constant $\kappa$
\cite{Stoks:1999bz}, we do not include the contribution from
exchanging $\Sigma$ baryon in Fig.~\ref{fig1}(c). There are also
other rescattering processes that can contribute to the decay
channel $J/\psi\to \gamma p\bar{p}$ as illustrated in
Fig.~\ref{fig1}(d). But note that the strong decay $J/\psi\to VP$
will exchange three gluons as a minimum number, while the production
of $\eta$ resonances in Fig.~\ref{fig1}(a)-(c) can occur via
exchanging two gluons. The transition of Fig.~\ref{fig1}(d) will be
relatively suppressed. Thus, we do not include their contribution at
this moment.

We distinguish contributions from light pseudoscalar meson such as
$\eta(547)$. Such states have relatively small couplings to
$J/\psi\gamma$ which can be determined by the $J/\psi$ radiative
decays. Also, we have better knowledge on their couplings to
nucleons. Since their masses are far below the $p\bar{p}$ threshold,
their contributions to $J/\psi\to \gamma p\bar{p}$ are strongly
suppressed. We call such contributions as direct couplings and they
are presented by Fig.~\ref{fig1}(e).

In the isoscalar channel for $p\bar{p}$, the large branching ratios
of $J/\psi\to \gamma \eta\to \gamma V\bar{V}$ and sizeable $VNN$
couplings actually allow us to study the form factors for
intermediate massive $\eta$ mesons to $p\bar{p}$ by $V\bar{V}$
rescatterings. Qualitatively, the $p\bar{p}$ invariant mass spectrum
could be sensitive to the dynamical details of the $\eta p\bar{p}$
form factors. This is different from treating the $\eta p\bar{p}$ by
a single coupling parameter. Our purpose is to explicitly calculate
the $\eta p\bar{p}$ form factors via intermediate $V\bar{V}$
rescatterings based on available experimental
data~\cite{Bugg:1995jq,Ablikim:2006ca,Bai:1999mk,Baltrusaitis:1985nd,Bisello:1988as}.

%\subsection{Effective Lagrangian}
The following effective Lagrangians are applied for the evaluation
of those Feynman diagrams:
\begin{eqnarray}
\mathcal {L}_{\rho NN} &=& g_{\rho NN}\bar{N}\left(\gamma^\mu
\vec{\tau}\cdot\vec{\rho}_\mu
+\frac{\kappa_\rho}{2m_N}\sigma^{\mu\nu}\vec{\tau}\cdot\partial_{\mu}\vec{\rho}_{\nu}
\right)N, \\
\mathcal{L}_{\omega NN} &=& g_{\omega
NN}\bar{N}\left(\gamma^{\mu}\omega_{\mu}+\frac{\kappa_\omega}{2m_N}\sigma^{\mu\nu}\partial_\mu
\omega_\nu \right)N, \\
\mathcal{L}_{K^* N\Lambda} &=& g_{K^*
N\Lambda}\bar{N}\left(\gamma^{\mu}\Lambda
K^*_{\mu}+\frac{\kappa_{K^*}}{2m_N}\sigma^{\mu\nu}\Lambda\partial_\mu
K^*_\nu \right)+H.c., \\
\mathcal{L}_{\gamma\psi\eta} &=&
e\frac{g_{\gamma\psi\eta}}{m_{\psi}}\epsilon_{\alpha\beta\gamma\delta}\partial^\alpha
A^\beta \partial^\gamma \psi^\delta \eta,\\
\mathcal{L}_{VV\eta} &=&
\frac{g_{VV\eta}}{m_{V}}\epsilon_{\alpha\beta\gamma\delta}\partial^\alpha
V^\beta \partial^\gamma \bar{V}^\delta \eta, \\
\mathcal {L}_{\eta NN} &=& -i g_{\eta NN} \bar{N}\gamma_{5} \eta N, \\
\mathcal{L}_{\omega\psi\eta} &=&
\frac{g_{\omega\psi\eta}}{m_{\psi}}\epsilon_{\alpha\beta\gamma\delta}\partial^\alpha
\psi^\beta \partial^\gamma \omega^\delta \eta,
\end{eqnarray}
where $\vec{\tau}$ are Pauli matrices, $\vec{\rho}$ denotes isospin
triplet, $N$ and $K^*$ denote isospin doublets which are defined as
follows:
\begin{eqnarray}
N=\left(
    \begin{array}{c}
      p \\
      n \\
    \end{array}
  \right),\ \ \
K^*=\left(
      \begin{array}{c}
        K^{*+} \\
        K^{*0}\\
      \end{array}
    \right),
\end{eqnarray}
and $\eta$ and $V$ denote the pseudoscalar and vector fields,
respectively. In our framework, $\eta$ represents $\eta(1405/1475)$,
$\eta(1760)$~\cite{Amsler:2008zzb} and a broad $0^-$ resonance
$X(1800)$~\cite{Bai:1999mk}, respectively.

The momenta of the intermediate meson rescatterings in
Fig.~\ref{fig1}(a)-(c) are denoted as
$J/\psi(P)\to\gamma(k)\eta(k_1)\to \gamma
V(q_1)\bar{V}(q_2)\to\gamma p(p_1)\bar{p}(p_2)$. Then the amplitude
is given by
\begin{eqnarray}
\mathcal{M}_{\eta} &=& e\frac{g_{\gamma\psi\eta}}{m_\psi}
\frac{\epsilon_{\alpha\beta\gamma\delta}k^\alpha
\epsilon^{*\beta}P^{\gamma}\epsilon_{\psi}^\delta}{s-m_{\eta}^2+im_\eta\Gamma_\eta}\int \frac{d^4q}{(2\pi)^4} \nonumber\\
&\times&  \frac{A(\eta\to VV\to
p\bar{p})}{(q_1^2-m_V^2)(q_2^2-m_V^2)}\mathcal{F}(q^2), \label{amp}
\end{eqnarray}
where
\begin{eqnarray}
&&A(\eta\to VV\to p\bar{p}) \nonumber\\ &\equiv&
\frac{g_{VV\eta}}{m_V}\epsilon_{\alpha\beta\gamma\delta}q_1^\alpha
q_2^\gamma \times g_{VBB}^2
\bar{u}(p_1)\left(\gamma^\beta+\frac{i\kappa}{2m_N}\sigma^{\beta\mu}q_{1\mu}\right) \nonumber \\
&&\times\frac{\rlap{/}{q}+m_B}{q^2-m_B^2}\left(\gamma^\delta+\frac{i\kappa}{2m_N}\sigma^{\delta\nu}q_{2\nu}\right)
v(p_2).
\end{eqnarray}
Since the exchanged baryon is off-shell, we introduce a dipole form
factor \cite{Cheng:2004ru} as follows to eliminate the divergence of
the loop-momentum integral:
\begin{eqnarray}
\mathcal{F}(q^2)=\left(\frac{\Lambda^2-m_B^2}{\Lambda^2-q^2}\right)^n,
\end{eqnarray}
with $n=2$. We note that a monopole form factor, i.e. $n=1$, will be
unable to kill the divergence. In principle, the experimental data
for $p\bar{p}$ annihilation into $VV$ can provide some constraints
on the form factors and couplings. We will discuss this later in
details. To evaluate the loop amplitude, we apply the software
package LoopTools \cite{Hahn:1998yk}.

\begin{table}
\begin{tabular}{|c|c|c|c|c|}
  \hline
   & $\rho NN$ & $\omega NN$ & $K^*N\Lambda$ &$K^*N\Sigma$ \\ \hline
  $g_{VBB}$ & $2.97$ & $10.36$ & $-4.26$ & $-2.46$ \\ \hline
  $\kappa$& $4.22$ & $0.41$ & $2.66$ &  $-0.47$\\
  \hline
\end{tabular}
\caption{Coupling constants of $VBB$ taken from
Ref.~\protect\cite{Stoks:1999bz}.}\label{table2}
\end{table}

The $VBB$ couplings are taken from the Nijmegen potential model
\cite{Stoks:1999bz}, and listed in Table~\ref{table2}. Although the
values may be different among different models, they are all within
a commonly accepted range in the literature. We define
\begin{eqnarray}
g_A\equiv e\frac{g_{\gamma\psi\eta}}{m_\psi}\frac{g_{VV\eta}}{m_V},
\end{eqnarray}
thus, the couplings $g_A$ can be determined by the branching ratios
$BR(J/\psi\to\gamma\eta)\times BR(\eta\to V\bar{V})$ listed in
Table~\ref{table1}. The numerical values are displayed in
Table~\ref{table3}.

\begin{table}
\begin{tabular}{|c|c|c|c|c|}
  \hline
   & $\eta(1405/1475)\rho\rho$ &$\eta(1760)\omega\omega$ &$\eta(1760)\rho\rho$& $(0^-)K^*\bar{K}^*$
   \\ \hline
  $g_A\ (\mbox{GeV}^{-2})$ & $0.024$ & $0.015$ & $0.007$ & $0.038$ \\
  \hline
\end{tabular}
\caption{Couplings of $J/\psi\to\gamma \eta\to \gamma V\bar{V}$ for
different intermediate pseudoscalar mesons.}\label{table3}
\end{table}

An interesting feature arising from the $\rho\rho$ and
$\omega\omega$ rescatterings is that all of them contribute to
$p\bar{p}$ significantly. Note that these three amplitudes have
absorptive part which can be determined in the on-shell
approximation. We find that they individually overestimate the
branching ratios for $J/\psi\to \gamma p\bar{p}$, and turn out to be
much larger than the direct transitions. This phenomenon suggests
that rescattering amplitudes of the intermediate
$\eta(1405/1475)\to\rho\rho$, $\eta(1760)\to\omega\omega$ and a
relatively smaller one $\eta(1760)\to\rho\rho$ should have a
destructive interference to suppress the overall amplitude in order
to be consistent with the experimental magnitude~\cite{Bai:2003sw}.
Because of this constraint, we introduce two relative phases
$e^{i\theta}$  and $e^{i\phi}$ between these amplitudes, i.e.
\begin{eqnarray}
\mathcal{M}=\mathcal{M}_{\eta}^{dir}+
\mathcal{M}_{\eta(1405)}^{\rho\rho,res}+e^{i\theta}\mathcal{M}_{\eta(1760)}^{\omega\omega,res}
+e^{i\phi}\mathcal{M}_{\eta(1760)}^{\rho\rho,res},
\end{eqnarray}
where $\mathcal{M}^{dir}$ and $\mathcal{M}^{res}$ denote the direct
and rescattering amplitudes, respectively. We note that the mesons
in the loops are generally treated as fundamental fields with
infinitely narrow widths in the effective Lagrangian approach. The
relative phase angles are thus introduced to take into account the
size effects arising from the meson propagators and interaction
vertices as commonly adopted. Since the contributions from the
direct transitions are negligibly small, the relative phases are not
sensitive to them and we do not discuss them in the following parts.
In comparison with the data~\cite{Bai:2003sw} we find that the
relative phases $\theta\simeq \pi$ and $\phi\simeq -\pi/2$ leading
to destructive interferences are favored.

\begin{figure}[tb]
\includegraphics[width=0.75\hsize]{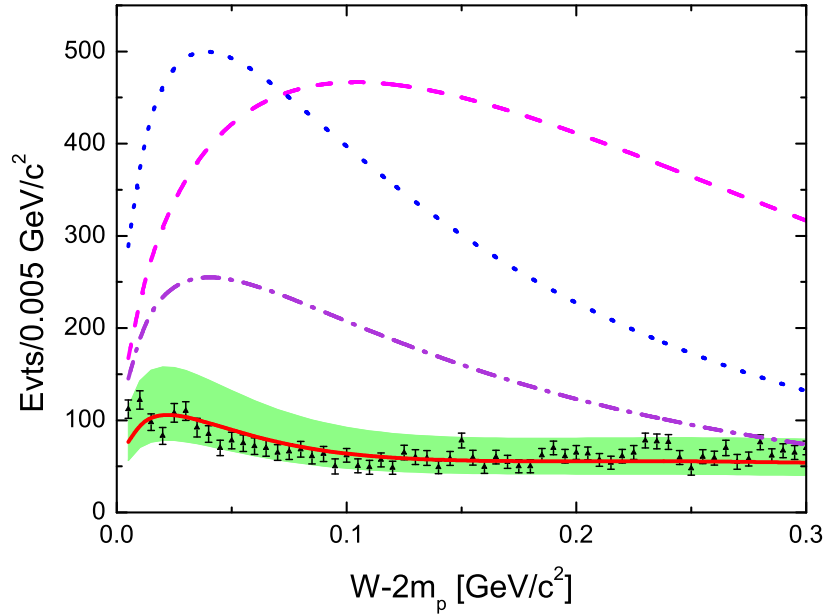}
\caption{The $p\bar{p}$ invariant mass spectrum of $J/\psi\to\gamma
p\bar{p}$. The dashed, dotted, and dot-dashed lines correspond to
contributions from $\rho\rho_{\eta(1405)}$,
$\omega\omega_{\eta(1760)}$, and
$\rho\rho_{\eta(1760)}$-rescattering, respectively. The solid line
is the overall interference with $\Lambda \simeq 1.17\ \mbox{GeV}$
and $\theta\simeq\pi$, $\phi\simeq -\pi/2$. The lower and upper
bound of the shadowed area correspond to $\Lambda=1.15$ and $1.20\
\mbox{GeV}$, respectively. The triangle with error bar represents
the experimental data from
Ref.~\protect\cite{Bai:2003sw}.}\label{fig2}
\end{figure}

\section{Numerical results and discussions}

\subsection{Results for $J/\psi\to \gamma p\bar{p}$}

In Fig.~\ref{fig2}, we plot the invariant mass spectrum of the
$p\bar{p}$ for the rescattering transitions, i.e.
$\eta(1405/1475)\to\rho\rho$, $\eta(1760)\to\omega\omega$ and
$\eta(1760)\to\rho\rho$, respectively. The coherent results are
shown by the solid curve, with $\Lambda\simeq 1.17\ \mbox{GeV}$, and
a bound given by $\Lambda=1.15\sim 1.20$ GeV. It shows that the
contributions from the $\eta(1405/1475)\to\rho\rho$,
$\eta(1760)\to\omega\omega$ and $\eta(1760)\to\rho\rho$
rescatterings are much larger than their coherent sum. Nevertheless,
at large value of $(W-2m_p)$, these contributions have different
behaviors. It is interesting to see the consequence of the
interferences among these amplitudes which produces the enhancement
at low $(W-2m_p)$ and flattened cross sections at large $(W-2m_p)$.

There are essential points which should be clarified here:

I) The $V\bar{V}$ rescattering mechanism can be recognized as a
dynamical account of the energy-dependent $\eta p\bar{p}$ form
factors.

II) We emphasize again that such a prescription is based on the
experimental evidence for the dominant $0^{-+}$ partial wave for
$V\bar{V}$ in $J/\psi\to \gamma V\bar{V}$, and significantly large
$V NN$ couplings. Therefore, we can expect to gain much better
insights into $J/\psi\to \gamma p\bar{p}$ reaction mechanism, in
particular, for the $0^{-+}$ partial wave in the $p\bar{p}$
spectrum.

III) The large cross sections given by intermediate $V\bar{V}$
rescatterings imply that there must exist destructive phases among
those amplitudes for which phase angles $\theta$ and $\phi$ are
introduced for the dominant transition amplitudes. This feature can
be regarded as less model-dependent since large contributions from
the $V\bar{V}$ rescatterings seem to be inevitable. We shall
investigate this in $p\bar{p}$ annihilation later to show that the
$V\bar{V}$ rescatterings have not been overestimated. In contrast,
we note that the behavior of the cancellations would be
model-dependent. The phase angles are determined in such a way that
the cancellations is required to produce the threshold enhancement
in the $p\bar{p}$ spectrum. However, for the purpose of exploring
possibilities of producing the threshold enhancement in $p\bar{p}$
spectrum, such a requirement can be justified.

IV) We note that the data in Ref.~\cite{Bai:2003sw} do not contain
sufficient background estimate as emphasized by BES~\cite{jin}. In
particular, the data contain contaminations from $\pi^0 p\bar{p}$,
and the detector efficiency (DE) has not been corrected. This will
affect the determination of $\Lambda$ which requires a better
understanding of those pieces of information. As shown by the dotted
curve in Fig. 3(a) of Ref.~\cite{Bai:2003sw}, the DE exhibits an
overall flattened shape though it is slightly better at small
$p\bar{p}$ invariant masses.  Because of this, a shadowed area
corresponding to $\Lambda=1.15\sim 1.20$ GeV is shown in
Fig.~\ref{fig2}. We can also see that the DE correction and
background subtraction will not change the shape of the enhancement
drastically.

\subsection{Results for $p\bar{p}\to V\bar{V}$}

As follows, we come to the key issue to investigate the $p\bar{p}\to
V\bar{V}$ as an independent check of the $V\bar{V}$ rescattering
mechanism. There are experimental data for $p\bar{p}$ annihilations
into vector meson
pairs~\cite{Amsler:1993kg,Baltay:1966zza,Klempt:2005pp,Dover:1992vj,Bruckner:1987ve}.
By adopting the same couplings used in $J/\psi\to \gamma p\bar{p}$
via $V\bar{V}$ rescatterings, we can calculate the cross sections
for $p\bar{p}\to V\bar{V}$ and then check whether the
$V\bar{V}$-rescattering contributions have been overestimated or
not.

The branching ratios for the $\omega\omega$ and $\rho^0\rho^0$ final
state in $p\bar{p}$ annihilations at rest were measured,
$BR(p\bar{p}\to \omega\omega)=(3.32\pm 0.34)\times 10^{-2}$
\cite{Amsler:1993kg} and $BR(p\bar{p}\to \rho^0\rho^0)=(0.4\pm
0.3)\times 10^{-2}$ \cite{Baltay:1966zza}. However, the total cross
sections with $p$ and $\bar{p}$ at rest are not available. We then
adopt the total cross section, $\sigma_T = 250\ \mbox{mb}$ with
$p_{Lab}= 200\ \mbox{MeV/c}$ for the incoming antiproton beam to
estimate the $\omega\omega$ and $\rho^0\rho^0$ production cross
sections. It gives
\begin{eqnarray}
\sigma_{exp}(p\bar{p}\to \omega\omega) &\approx& 8.3 \pm  0.85\
\mbox{mb},
\nonumber \\
\sigma_{exp}(p\bar{p}\to \rho^0\rho^0)&\approx& 1 \pm  0.75\
\mbox{mb}. \nonumber
\end{eqnarray}
For $p\bar{p}$ with low relative momenta, the cross sections should
be dominated by the relative $S$ wave, i.e. the orbital angular
momentum between $p$ and $\bar{p}$ is zero. Furthermore, the
configuration of $^{2S+1}L_J= \ ^3S_1$ will have $C=-1$. Thus, it
will be suppressed due to $C$-parity violation when it couples to
$\omega\omega$ and $\rho^0\rho^0$. The $S$-wave decay will then
occur via the $^1S_0$ configuration, and the cross section can be
estimated by using the following projector for the $p\bar{p}$ system
\cite{Bodwin:2002hg}:
\begin{eqnarray}
\Pi_0(p_1,p_2)&=&-\sum\limits_{\lambda_1,\lambda_2}
u(p_1,\lambda_1)\bar{v}(p_2,\lambda_2)\langle\frac{1}{2}\lambda_1\frac{1}{2}\lambda_2|00\rangle
\nonumber \\
&=&\frac{1}{2\sqrt{2}(E+m_N)}(\rlap{/}{p}_1 +m_N)(1+\gamma^0)
\gamma_5 (\rlap{/}{p}_2 -m_N),
\end{eqnarray}
where $E=\sqrt{(p_1+p_2)^2}/2$.

With the same effective Lagrangians, form factors and coupling
constants as in $J/\psi\to \gamma p\bar{p}$, and with the cut-off
energy $\Lambda=1.15 \sim 1.20\ \mbox{GeV}$, we obtain the following
cross sections:
\begin{eqnarray}
\sigma_{th}(p\bar{p}\to \omega\omega) &\approx&  2.4\sim 4.7\
\mbox{mb},
\nonumber \\
\sigma_{th}(p\bar{p}\to \rho^0\rho^0)&\approx& 2.0 \sim 3.9\
\mbox{mb}, \nonumber
\end{eqnarray}
which are consistent with the data within both experimental and
theoretical uncertainties. This suggests that our $V\bar{V}$
rescattering contributions have not been overestimated. We note that
the full calculation without imposing the $^1S_0$ projector gives
similar results near threshold which confirms the $S$-wave
dominance.

\begin{figure}[tb]
\includegraphics[width=0.45\hsize]{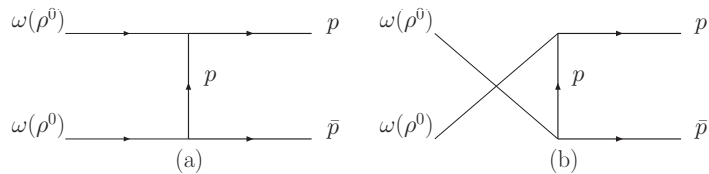}
\caption{Diagrams for the process $V\bar{V}\to p\bar{p}$.
}\label{fig3}
\end{figure}

\begin{figure}[tb]
\includegraphics[width=0.75\hsize]{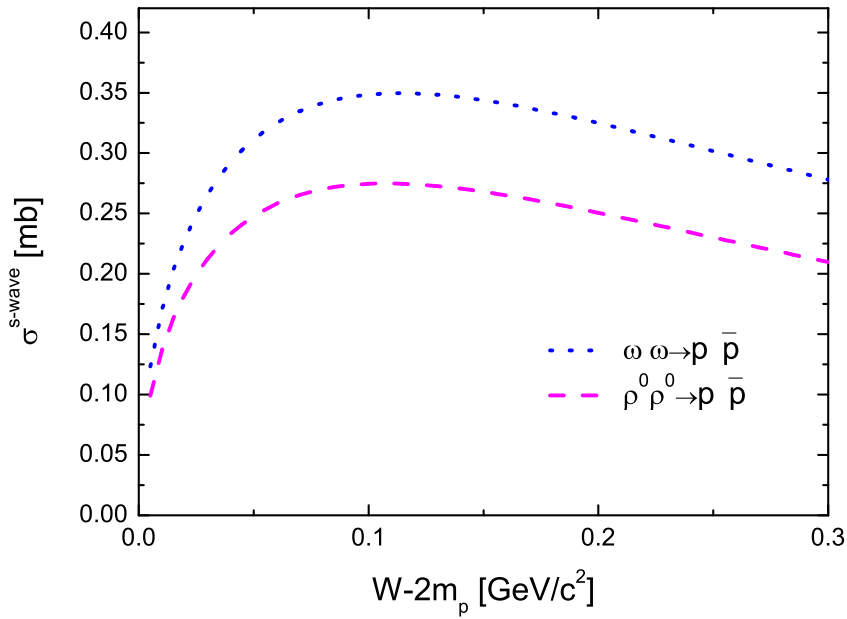}
\caption{Energy dependence of cross sections for $V\bar{V}\to
p\bar{p}$ with $p\bar{p}$ in the $^1 S_0$ state.}\label{fig4}
\end{figure}

We can further understand the $V\bar{V}$ rescattering mechanism by
looking at the $V\bar{V}\to p\bar{p}$. The reaction can be
illustrated by Feynman diagrams in Fig.~\ref{fig3}. The $^1S_0$
configuration is also dominant near threshold, i.e. an $S$-wave
decay amplitude. Again, with the same effective Lagrangians, form
factors and couplings, we find a quick increase of the cross
sections at small values of $(W-2m_p)$ as shown in  Fig.~\ref{fig4}.
It helps clarify that the intermediate $V\bar{V}$ rescatterings can
contribute to the threshold enhancement in the $p\bar{p}$ invariant
spectrum in $J/\psi\to \gamma p\bar{p}$.

\subsection{Results for $J/\psi\to \omega p\bar{p}$}

\begin{figure}[tb]
\includegraphics[width=0.75\hsize]{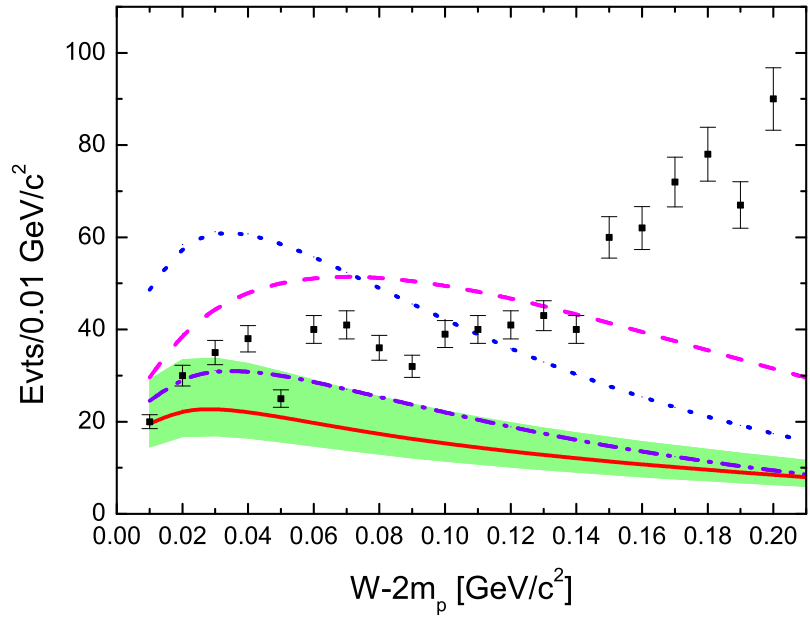}
\caption{The $p\bar{p}$ invariant mass spectrum of $J/\psi\to\omega
p\bar{p}$. The notation is the same as Fig. \protect\ref{fig2}, and
the experimental data are from Ref.~\protect\cite{:2007dy}.
}\label{fig5}
\end{figure}

Further search for this threshold enhancement was carried out in
$J/\psi\to \omega p\bar{p}$ at BES~\cite{:2007dy}, where it was
claimed that the $p\bar{p}$ enhancement was absent as shown by the
data in Fig.~\ref{fig5}. However, Haidenbauer {\it et
al.}~\cite{Haidenbauer:2008qv} suggest that there still exist a
similar threshold enhancement due to final state $p\bar{p}$
interaction except that it is much less significant due to kinematic
changes and competing hadronic background.

We also extend our formalism to $J/\psi\to\omega p\bar{p}$. The
measurement of $J/\psi\to \omega\eta(1405)$ by BES
\cite{Ablikim:2007ev} allows us to estimate,
\begin{equation}
\mbox{BR}(J/\psi\to\omega \eta(1405))\sim 10^{-3},
\end{equation}
with which the coupling constant is extracted as
\begin{equation}
\frac{g_{\omega\psi\eta}}{m_\psi}\frac{g_{\rho\rho\eta}}{m_\rho}
\simeq 0.0093\ \mbox{GeV}^{-2}.
\end{equation}
With the other parameters fixed the same as in $J/\psi\to \gamma
p\bar{p}$, we plot the $p\bar{p}$ invariant mass distribution for
$J/\psi\to\omega p\bar{p}$ in Fig.~\ref{fig5}. Again, it shows that
the rescattering terms overestimate the cross sections at low
$(W-2m_p)$, while the interference gives much smaller cross section.
We do not try to quantitatively describe the data at large invariant
masses since they are the kinematics that other partial waves and
mechanisms would become important.

It should be noted that our estimate of the $\omega p\bar{p}$ decay
is rather rough, and a better measurement of
$\mbox{BR}(J/\psi\to\omega \eta(1405))$ and
$\mbox{BR}(J/\psi\to\omega \eta(1760))$ will provide a better
constraint on our model. However, this does not prevent us from
gaining some insights into the $p\bar{p}$ threshold enhancement due
to final state interactions. It is essential to recognize that the
rescattering transitions via $V\bar{V}$ could be much larger than
the direct transitions based on the available experimental
evidence~\cite{Amsler:2008zzb} and the significant absorptive
contributions from the $V\bar{V}$ rescatterings. This can be
regarded as a peculiar property of some of those $\eta N\bar{N}$
off-shell couplings. Additional experimental information from
$p\bar{p}$ annihilations seems to confirm such a dynamics. Although
the determination of the relative phases depends on the requirement
of cancellations among the dominant amplitudes, we emphasize that
the presence of the threshold enhancement is mainly due to the
property of $V\bar{V}\to p\bar{p}$ transitions.

\section{Summary}
It is of great importance to recognize that the same mechanism may
behave differently in different channels due to kinematic and
interferences from other processes. Therefore, it may not appear
prominently everywhere. Because of this, it appears to be an
attractive solution for our understanding of the $J/\psi\to \gamma
p\bar{p}$ and $\omega p\bar{p}$ results. As studied in the
literature~\cite{Zou:2003zn,Kerbikov:2004gs,Bugg:2004rk,Sibirtsev:2004id,Entem:2007bb,Chen:2008ee,Haidenbauer:2008qv}
that $p\bar{p}$ final state interaction can also produce threshold
enhancement, it is urged to have a systematic understanding of how
these mechanisms exhibit and interfere with each other. We expect
that the BES-III experiment in the near future would provide a great
opportunity to clarify the underlying dynamics of the $p\bar{p}$
threshold enhancement~\cite{Asner:2008nq}.

\section*{Acknowledgement}

Authors thanks B.S. Zou, S. Jin and X.-Y. Shen for useful
discussions. This work is supported, in part, by the National
Natural Science Foundation of China (Grants No. 10675131 and
10491306), Chinese Academy of Sciences (KJCX3-SYW-N2), and the
Ministry of Science and Technology of China (2009CB825200).

%\newpage

\end{document}